\documentclass[twocolumn, showpacs,superscriptaddress]{revtex4}
\usepackage{epsfig}
\usepackage{amssymb}
\usepackage{amsmath}
\usepackage{amsbsy}
\usepackage{color}
\usepackage{ulem}
\usepackage{epstopdf}

\begin{document}
\title{Nano-pore based characterization of branched polymers}
\author{Takahiro Sakaue}
\email{sakaue@phys.kyushu-u.ac.jp}
\affiliation{Department of Physics, Kyushu University 33, Fukuoka 812-8581, Japan}
\author{Fran\c{c}oise Brochard-Wyart}
\email{francoise.brochard@curie.fr}
\affiliation{UPMC Univ Paris 06, CNRS UMR 168,Institut Curie, 26 rue d'Ulm, 75248 Paris Cedex 05, France}

\begin{abstract}
We propose a novel characterization method of randomly branched polymers based on the geometrical property of such objects in confined spaces. The central idea is that randomly branched polymers exhibit passing/clogging transition across the nano-channel  as a function of the channel size. This critical channel size depends on the degree of the branching, whereby allowing the extraction of the branching information of the molecule.
\end{abstract}


\maketitle

\section{Introduction}
\label{Intro}
The concept of the confinement plays an important role in various aspects of soft matter research ranging from diverse industrial applications to our fundamental understanding of soft materials. 
A linear polymer in confined spaces is a classical example, which provides one of the best opportunities to illustrate the scaling concept in polymer physics~\cite{deGennes}. 
In the present paper, we turn our attention to more complex object, {\it randomly branched polymers}, under confinement. The source of the complexity arises from their nontrivial, quenched connectivity, which introduces some unexpected features, compared to the linear polymer counterpart, in narrow channel.

Some time ago, we analyzed the injection process of such branched object into a narrow channel by applying fluid flow~\cite{Flow-injection, Sakaue_Raphael}.  If the critical fluid current to achieve the injection depends on the molecular parameters, such as the molecular weight and the degree of the branching, one can characterize the molecule from the measurement. However, it turned out that the critical current depends neither on the molecular parameters nor the channel size. In this paper, we come back to our original motivation and propose an alternative characterization method using a finite-length channel. In Sec.~\ref{Sec:2}, we summarize basic static and dynamical properties of randomly branched polymers confined in channel following earlier works by emphasizing their unique properties in comparison to linear polymers~\cite{Flow-injection, Sakaue_Raphael, Vilgis, Gay, deGennes_pore}. The scaling formulae for the free energy of confinement and diffusion coefficient are given, which are expressed in terms of fundamental length scales in the problem. In Sec.~\ref{Partial}, we look at the confinement process of the branched object and point out that it is a progressive process. We then propose a characterization method in Sec.~\ref{Characterization}.

\section{Branched polymer in narrow channel}
\label{Sec:2}
Let us recall the conformational properties of randomly branched polymers.
The spatial size, i.e., the radius of gyration, of the ideal branched polymer with $N$ monomers of size $a$ is $R_0 \simeq a b^{1/2}(N/b)^{1/4} \simeq a N^{1/4}b^{1/4}$, where $b$ is the average number of monomers between consecutive branching points~\cite{Zimm_Stockmayer, deGennes_Branch}.
In good solvent, the swelling occurs due to the excluded-volume effect, which can be evaluated from the following  Flory-type free energy~\cite{Isaacson,Daoud_Joanny}
\begin{eqnarray}
\frac{F(R)}{k_BT} \simeq \frac{R^2}{R_0^2} + \frac{N^2 a^3}{R^3}
\label{eq:1}
\end{eqnarray}
where $k_BT$ is the thermal energy.
Minimization of above free energy with respect to $R$ leads the equilibrium size 
\begin{eqnarray}
R \simeq a N^{1/2}b^{1/10} \simeq a b^{3/5}\left( \frac{N}{b}\right)^{1/2}
\label{R_bulk}
\end{eqnarray}
One can check the crossover to the linear chain behavior $R\rightarrow a N^{3/5}$ in the limit $b \rightarrow N$.

Now, let us confine the branched polymer into a narrow channel with the diameter $D \ll R$.
The polymer is stretched along the channel axis with the length $R_{\parallel} > R$. Noting that the available volume becomes $\sim D^2 R_{\parallel}$, Flory free energy is then modified as
\begin{eqnarray}
\frac{F(R_{\parallel})}{k_BT} \simeq \frac{R_{\parallel}^2}{R_0^2} + \frac{N^2 a^3}{D^2 R_{\parallel}}
\label{eq:2}
\end{eqnarray}
which yields the optimum extension as
\begin{eqnarray}
R_{\parallel} \simeq a \left( \frac{a}{D}\right)^{2/3}N^{5/6}b^{1/6}
\label{eq:R_parallel}
\end{eqnarray}
From the condition of the highest possible packing, $\phi \simeq 1$
with the volume fraction
\begin{eqnarray}
\phi \simeq Na^3/(D^2 R_{\parallel}) \simeq (a/D)^{4/3}(N/b)^{1/6} 
\label{eq:phi}
\end{eqnarray}
we find the {\it minimum channel size}~\cite{Vilgis}
\begin{eqnarray}
D_{min} = a \left( \frac{N}{b}\right)^{1/8}c^{-3/4}
\label{eq:D_min}
\end{eqnarray}
Here, we have purposely introduced the numerical coefficient $c$ through the definition $Na^3/(D_{min}^2 R_{\parallel}) = c$ for $D_{min}$, which would be useful for the concrete evaluation of the numerical values in Sec.~\ref{Characterization}.
The branched polymer can not be fully squeezed into a narrower channel with $D < D_{min}$. 
The corresponding extension $L_A $ is what we call the {\it Ariadne length}~\cite{Gay}
\begin{eqnarray}
L_A =  R_{\parallel}(D_{min}) = a b \left(\frac{N}{b}\right)^{3/4}c^{1/2}
\label{eq:L_A}
\end{eqnarray}
Note that Eq.~(\ref{eq:R_parallel}) can be arranged into the form
\begin{eqnarray}
R_{\parallel} \simeq L_A \left(\frac{D_{min}}{D} \right)^{2/3}
\label{eq:R_parallel_2}
\end{eqnarray}
As this expression shows, $L_A$ and $D_{min}$ are basic length scales dictating the scaling properties of the confined branched objects.

\subsection{Free energy for the confinement}
\label{FreeEnergy}
The Ariadne length corresponds to the maximum chemical distance along the internal coordinate of the connectivity. 
In the above argument, we realize the maximum stretching of branched polymer by confining it into the narrowest channel.
Therefore, rewriting Eq.~(\ref{eq:L_A}) into the form
\begin{eqnarray}
\frac{N}{b} \simeq \left( \frac{L_A}{ab}\right)^{d_s}
\end{eqnarray}
indicates that our argument identifies the branched object to be a polymeric fractal with the spectral dimension $d_s = 4/3$.

Since the embedding space dimension $d=1$ in the channel geometry is lower than the spectral dimension, we encounter what we call the {\it strong confinement} regime~\cite{Sakaue_Raphael, Sakaue_Semiflexible}~\footnote{If we make a similar calculation for a slit  (d=2 larger than ds=4/3), we are not in the strong confinement regime. It would be interesting to compare the two cases.}. The equivalent statement is that the space dimension of the channel geometry ($d=1$) is below the lower critical dimension $d_c = d_s$ for the object~\cite{Cates}. In such a situation, it would be natural to introduce a length scale $\xi$ from the closed packed condition of blobs (with $g$ segments in each) $a^3 g/\xi^3 \simeq \phi$. Combining it with Eq.~(\ref{eq:phi}), we find
\begin{eqnarray}
\xi &\simeq& a b^{-1/5}\left( \frac{a}{D}\right)^{-4/3}\left( \frac{N}{b}\right)^{-1/6} \nonumber \\
&\simeq& a \left( \frac{D_{min}}{D^*}\right)^{1/3}\left( \frac{D}{D_{min}}\right)^{4/3}  \qquad (D > D^*)
\label{xi_N_1}
\end{eqnarray}
where, as usual for the semidilute solution, the bulk statistic (Eq.~(\ref{R_bulk}))
\begin{eqnarray}
\xi \simeq a g^{1/2}b^{1/10}
\label{xi_1}
\end{eqnarray}
  is assumed inside the blob~\cite{deGennes}. 
Upon decreasing $D$, Eq.~(\ref{xi_1}), and hence, Eq.~(\ref{xi_N_1}), too, cease to be valid at $g \simeq b \Leftrightarrow D \simeq D^{*} \simeq b^{3/5} D_{min}$, where the blob becomes small enough so that it only probes the linear structure between the consecutive branching points~\cite{Gay, deGennes_pore}. Then Eq.~(\ref{xi_1}) should be modified to the linear chain statistics
\begin{eqnarray}
\xi \simeq a g^{3/5}
\label{xi_2}
\end{eqnarray}
so that
\begin{eqnarray}
\xi \simeq D \left( \frac{N}{b}\right)^{-1/8} \simeq a \left( \frac{D}{D_{min}}\right) \qquad (D < D^{*})
\label{xi_N_2}
\end{eqnarray}

 On applying $k_BT$ per blob ansatz, the free energy $\Delta F$ for confining the branched object can be obtained as
\begin{eqnarray}
\frac{\Delta F}{Nk_BT} \simeq
\left\{
           \begin{array}{ll}
    b^{3/5} \left( \frac{a}{D}\right)^{8/3}\left( \frac{N}{b}\right)^{1/3}  \simeq  \left( \frac{D^*}{D_{min}}\right) \left( \frac{D_{min}}{D}\right)^{8/3}         & \\
   \qquad \qquad \qquad \qquad ({\rm for} \ D > D^*) &  \\
   \\
    
    \left( \frac{a}{D}\right)^{5/3}\left( \frac{N}{b}\right)^{5/24} \simeq  \left( \frac{D_{min}}{D}\right)^{5/3}            . & \\
  \qquad \qquad \qquad \qquad ({\rm for} \ D < D^*)  &
           \end{array}
        \right.&&
        \label{D_F}
\end{eqnarray}
One can check $\Delta F(D^*) \simeq (N/b) k_BT$ and $\Delta F(D_{min}) \simeq N k_BT$.
Notice that $\Delta F$ increases faster than linear with $N$ at fixed $D$, which is a general feature in the strong confinement regime. The generalized version with the spectral dimension $d_s$ is presented in Ref.~\cite{Sakaue_Raphael}~\footnote{There are several misprints in the scaling exponent in Ref.~\cite{Sakaue_Raphael} Secs. III and IV, which are corrected in the footnote (11) in Ref.~\cite{Sakaue_Semiflexible}}.

\subsection{Diffusion coefficient}
\label{Diffusion}
Our blob picture indicates the following dynamical property of the confined branched object. As usual in the semidilute solution, we assume that the hydrodynamic screening length also scales as $\xi$ so that each blob contributes the Stokes friction $\sim \eta_0 \xi$ to its translational motion~\cite{deGennes}. Using Eq.~(\ref{xi_1})/~(\ref{xi_2}), we obtain the scaling formulae for the diffusion coefficient $\mathcal D$ of the confined branched object in channel
\begin{eqnarray}
{\mathcal D} \times \frac{\eta_0 a N}{k_BT} \simeq
\left\{
           \begin{array}{ll}
     b^{-2/5}\left( \frac{D}{a}\right)^{4/3}\left( \frac{b}{N}\right)^{1/6}  \simeq   \left( \frac{D_{min}}{D^*}\right)^{2/3}  \left( \frac{D}{D_{min}}\right)^{4/3}         & \\
   \qquad \qquad \qquad \qquad ({\rm for} \ D > D^*) &  \\
   \\
    
   \left( \frac{D}{a}\right)^{2/3}\left( \frac{b}{N}\right)^{1/12}  \simeq     \left( \frac{D}{D_{min}}\right)^{2/3}             . & \\
  \qquad \qquad \qquad \qquad ({\rm for} \ D < D^*)  &
           \end{array}
        \right.&&
        \label{D_D}
\end{eqnarray}

Notice that setting $b=N$, we recover the classical results for linear polymers $R_{\parallel} \simeq a (a/D)^{2/3}N$, $D_{min} \simeq a$, $L_A \simeq aN$, $D^* \simeq a N^{3/5}$, $\xi \simeq D$, $\Delta F/k_BT \simeq (a/D)^{5/3} N$ and ${\mathcal D} \simeq(k_BT/\eta_0 a N)(D/a)^{2/3}$.

\section{Partial confinement}
\label{Partial}
\begin{figure}[h]
\includegraphics[width=0.4\textwidth]{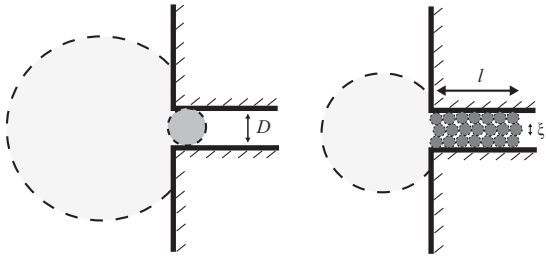}
\caption{Sketches of the partially confined branched object. The density inside the channel becomes higher with the process of the confinement.}
\label{Fig1}
\end{figure}

Next, we consider the process of the confinement: only a part of the whole chain with $n \ (<N)$  monomers is confined in the channel. The extension of the partially confined part with $n$ monomers in the channel is denoted as $l \ (<R_{\parallel})$ (Fig.~\ref{Fig1} ).

From Eq.~(\ref{eq:R_parallel}) with the replacement $(R_{\parallel}, N) \rightarrow (l, n)$, we find the number $n$ of monomers in confinement as a function of $D$ and $l$;
\begin{eqnarray}
n(D, l) \simeq \left( \frac{l}{a}\right)^{6/5}\left( \frac{D}{a}\right)^{4/5}b^{-1/5}
\label{eq:n_l}
\end{eqnarray}
The volume fraction of this partially confined region is
\begin{eqnarray}
\phi (D, l) \simeq \frac{n(D, l) a^3}{D^2 l} \simeq \left( \frac{l}{a}\right)^{1/5}\left(\frac{a}{D} \right)^{6/5}b^{-1/5}
\label{eq: phi_l}
\end{eqnarray}
As we have already argued in ref.~\cite{Flow-injection, Sakaue_Raphael}, the confinement is a progressive process, i.e., $\phi$ increases with $l$, which is again a hallmark of the strong confinement regime. In the limit of the complete confinement $l \rightarrow R_{\parallel}$, we recover Eq.~(\ref{eq:phi}).

In Refs.~\cite{Flow-injection, Sakaue_Raphael}, we have analyzed the fluid flow (or pressure drop) driven suction process of the branched object into the channel, where two competing factors, i.e., the osmotic force due to the confinement and the hydrodynamic drag force, are calculated in line with the argument in Secs.~(\ref{FreeEnergy}) and~(\ref{Diffusion}), respectively.
The main result is that the critical flow current $J_c$ for the injection is independent of both $N$ and $D$, and its scaling structure is given by
\begin{eqnarray}
J_c \simeq \frac{k_BT}{\eta_0}
\label{Jc}
\end{eqnarray}
which is the same as that for the linear chain~\cite{deGennes_pore, Beguin, Yeomans}.
Underlying physics for this counter-intuitive result is discussed in Ref.~\cite{Flow-injection, Sakaue_Raphael}.

\section{Confinement in finite-length channel}
\label{FiniteChannel}
From a practical viewpoint, the above result Eq.~(\ref{Jc}) on the critical injection current is disappointing, since it indicates no utility to characterize the probed branched molecules.
This has led us to propose an alternative characterization method using the {\it finite length} channel.
Now consider the confinement of branched polymers into a narrow channel with the length $L$. We ask the following question: Given the channel geometry $D$ and $L$, can the branched polymer with molecular architecture $N$ and $b$ pass through the channel? Here, the minimum size $D_{mim}$ and the Ariadne length $L_A$ come into the problem as relevant length scales.

The critical channel length $L_c$ can be derived from the condition $\phi(D, L_c) = c$ as
\begin{eqnarray}
L_c = a b \left( \frac{D}{a}\right)^{6}c^5 = L_A \left( \frac{D}{D_{min}}\right)^{6}
\label{eq:L_c_1}
\end{eqnarray}

\begin{figure}[h]
\includegraphics[width=0.4\textwidth]{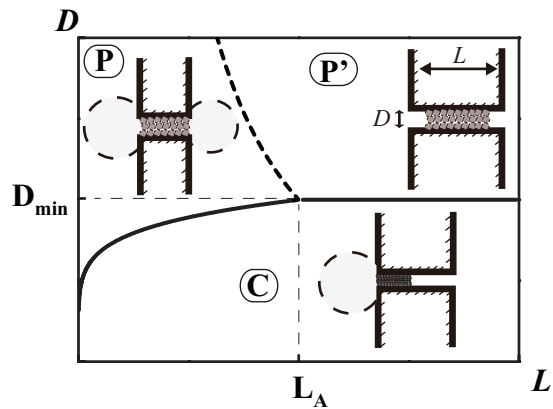}
\caption{Diagram of passing-clogging transition. The border between regimes P and C is given by Eq.~(\ref{eq:L_c_1}). The dashed curve between regimes P and P' is given by Eq.~(\ref{eq:R_parallel_2}).}
\label{Fig2}
\end{figure}
From this, we can construct the following diagram (Fig.~\ref{Fig2}).
In regimes P/P', the branched object can pass through the channel. The regime P and P' correspond to the short and long channels, respectively; in the latter, the complete confinement is realized. In the regime C, however, it cannot pass; the dense pack limit $\phi \rightarrow c$ is reached during the confinement process, which results in the molecular clogging.

To see the origin of the passing/clogging transition, it is instructive to compare the above results with the linear chain counterparts. By setting $b \simeq n$ in Eqs.~(\ref{eq:n_l}) and~(\ref{eq: phi_l}), one finds $n(D,l) \simeq (l/a)(D/a)^{2/3}$ and $\phi(D,l) \simeq (a/D)^{4/3}$.  Because of the independence of $\phi$ on $l$, there is no critical length $L_c$ for the passage. The filtration border is just given by the trivial geometrical condition $D = D_{min} \simeq a$.
In contrast, branched polymers allow one to exploit their unique property in channels to probe their molecular characteristics, the principle of which is outlined below.

\subsection{Characterization}
\label{Characterization}
Assume that we have a solution of branched polymers, which is mono-dispersed, but with unknown molecular parameters $N$ and $b$.  We would like to extract these parameters, in particular, $b$ the branching information. 

The measurement can be performed by fixing either (i) channel cross-sectional size $D$ or (ii)channel length $L$.
Let us first consider the case (i). For sufficiently short channel, the molecules can pass through it (regime P). Upon increasing the channel length, the molecular clogging may occur depending on the channel size, i.e., the transition to the regime C occurs at $L=L_c$, if $D < D_{min}$. 
By inverting the relation Eq.~(\ref{eq:L_c_1}) and using experimentally determined threshold value $L_c$ , we can deduce
\begin{eqnarray}
b = \frac{L_c}{a}\left( \frac{a}{D}\right)^{6 }c^{-5}
\label{eq:b_L}
\end{eqnarray}
where $c \simeq 0.5$ is a numerical coefficient (see the comment below Eq.~(\ref{eq:D_min})).
On the other hand, if the channel is not narrow enough, we do not expect the clogging; upon increasing $L$, we just enter the regime P', where the entire molecule can be confined during the passage.

In the case (ii), we fix the channel length $L$, and allow the channel cross-sectional size $D$ to vary. Such a tunable control is possible in soft elastomeric channel by applying the mechanical compressional force~\cite{elastomeric_channel}.
Since the channel length $L$ is fixed, it is convenient to rewrite Eq.~(\ref{eq:L_c_1}) as the threshold condition for the channel cross-sectional size:
\begin{eqnarray}
D_c = a \left( \frac{L}{ab}\right)^{1/6}c^{-5/6} = D_{min} \left( \frac{L}{L_A}\right)^{1/6}
\label{eq:D_c_1}
\end{eqnarray}

Upon decreasing $D$, we expect the transition from the passing regime P/P' to the clogging regime C.
If $N$ is large enough and we choose the short length channel in such a way as $L < L_A$ (see Eq.~(\ref{eq:L_A})), the threshold channel size $D_c$ is given by Eq.~(\ref{eq:D_c_1}).
By inverting the relation Eq.~(\ref{eq:D_c_1}) and using experimentally determined threshold value $D_c$ , we can deduce
\begin{eqnarray}
b = \frac{L}{a}\left( \frac{a}{D_c}\right)^{6 }c^{-5}
\label{eq:b}
\end{eqnarray}
On the other hand, for sufficiently long channels $L > L_A$, the threshold is given by the minimum diameter $D_c=D_{min}$ (eq.~(\ref{eq:D_min})).
Combining the results of measurements in two distinct transition regimes, i.e., (P $\rightarrow$ C) and (P' $\rightarrow$ C), one can obtain {\it both} molecular parameters $b$ and $N$.


From Eqs.~(\ref{eq:D_min}) and~(\ref{eq:L_A}), one can estimate (with $c \sim 0.5$), for instance, $D_{min} \sim 5a, \ L_A \sim 7000 a$ for $N \sim 10^5, \ b \sim 10$, and $D_{min} \sim 3a, \ L_A \sim 200 a$ for $N \sim 10^3, \ b \sim 5$. The very weak dependence of $D_{min}$ on molecular parameters (Eq.~(\ref{eq:D_min})) may be lucky; it enables one to cover a wide range of molecular parameters by a slight tuning of the channel diameter, indicating the feasibility of the protocol (ii).
For the efficient operation, one needs to apply the driving force (pressure gradient or some other means) above the threshold given by Eq.~(\ref{Jc}), which ensures the smooth entry into the channel.

\if
{\it Order estimation}---
$(N \sim 10^5, \ b \sim 10, \ c \sim 0.5) \rightarrow (D_{min} \sim 5a, \ L_A \sim 7000 a)$.

$(N \sim 10^4, \ b \sim 10, \ c \sim 0.5) \rightarrow (D_{min} \sim 4a, \ L_A \sim 1000 a)$.

$(N \sim 10^3, \ b \sim 5, \ c \sim 0.5) \rightarrow (D_{min} \sim 3a, \ L_A \sim 200 a)$.
\fi

\section{Summary}
Despite a multitude of research activity on linear polymers in confined spaces, its extension to more complex polymers is not abundant yet. In this paper, we have summarized the static and dynamical scaling properties of randomly branched polymers confined in channel. 
These are more complex than the linear polymer counterparts due to their quenched fractal connectivity characterized by the nontrivial spectral dimension $d_s = 4/3$~\footnote{There is also another class of the randomly branched polymer, in which the branching points fluctuate and rearrange, i.e., annealed branched polymer~\cite{Annealed}}. We repeat once more that the facts $D_{min} \gg a$ and $L_A \ll aN$ are the manifestation of the nontrivial connectivity, and these quantities $D_{min}$ and $L_A$ as basic length scales are indeed two sides of the same coin (Eq.~(\ref{eq:L_A})).

Emphasis has been put on their peculiar properties arising from the fact that the channel geometry is below their lower critical dimension, i.e., we are dealing with the strong confinement regime. We have provided the scaling formulae for the confinement free energy (Eq.~(\ref{D_F})) and diffusion coefficients (Eq.~(\ref{D_D})) , demonstrating their nontrivial dependence on $N$, and their compact expressions in terms of $D_{min}$.
There are some recent numerical~\cite{Likos} and experimental~\cite{Hyperbranch} attempts on related issues. Here, we once more repeat that the (statistical) connectivity pattern is crucial, according to which the molecules should be properly classified. A principle connectivity feature of the randomly branched polymer lies in its tree-like structure, i.e., no internal loop (cycle), and more quantitatively, as we have already seen, it is characterized by the spectral dimension $d_s = 4/3$. If, instead, the non-negligible fraction of the loops are formed during the synthesis, the resultant branched objects are more like the cross-linked micro- or nano-gels, whose elastic modulus is much higher than the randomly branched polymer, and is controlled by the mesh size. We thus expect that the critical injection current of such cross-linked objects into long channels increases with the molecular weight ($N$) and decreases with the subchain length ($b$), which may explain the experimental result in Ref.~(\cite{Hyperbranch}). It is clear that the dendrimers, studied in Ref.~(\cite{Likos}), are also quite different from our branched polymers. We hope that the current discussion would activate interest in this fascinating field of the complex polymers in confined spaces.

Finally, by exploiting exotic (but fundamental) properties of the branched objects in narrow channels, we have proposed a novel method for their characterization using the finite length channel. We hope our prediction and the feasibility of the method to be tested in near future.

"This work was supported by the JSPS Core-to-core Program, ``Non-equilibrium dynamics of soft matter and information'', and JSPS KAKENHI Grant Number 24340100, and the Labex CelTisPhyBio, N‹ANR-10-LBX-0038 part of the Idex PSL N‹ANR-IDEX-0001-02 PSL Paris.


\begin{thebibliography}{40}
\bibitem{deGennes}P.-G. de Gennes, {\it Scaling Concepts in Polymer Physics} (Cornell University Press, Ithaca, 1979).
\bibitem{Flow-injection}T. Sakaue, E. Rapha\"el, P.-G. de Gennes and F. Brochard-Wyart, Europhys. Lett. {\bf 72}, 83 (2005)
\bibitem{Sakaue_Raphael}T. Sakaue and E. Rapha\"el, Macromolecules {\bf 39}, 2621 (2006).
\bibitem{Vilgis}T.A. Vilgis, P. Haronska and M. Benhamou, J. Phys. II, {\bf 4}, 2187 (1994).
\bibitem{Gay}C. Gay, P.-G. de Gennes, E. Rapha\"el and F. Brochard-Wyart, Macromolecules {\bf 29}, 8379 (1996).
\bibitem{deGennes_pore}P.-G. de Gennes, Adv. Polym. Sci., {\bf 138}, 91 (1999).
\bibitem{Zimm_Stockmayer}B.H. Zimm and W.H. Stockmayer, J. Chem. Phys. {\bf 17}, 1301 (1949).
\bibitem{deGennes_Branch}P.-G. de Gennes, Biopolymers {\bf 6}, 715 (1968).
\bibitem{Isaacson}J. Isaacson and T.C. Lubensky, J. Phys. Lett. {\bf 41}, L649 (1980).
\bibitem{Daoud_Joanny}M. Daoud, J.F. Joanny, J. Phys. France {\bf 42}, 1359 (1981).
\bibitem{Sakaue_Semiflexible}T. Sakaue, Macromolecules {\bf 40}, 5206 (2007).
\bibitem{Cates}M.E. Cates, Phys. Rev. Lett., {\bf 53}, 926 (1984).
\bibitem{Beguin}L. B\'eguin, B. Grassl, F. Brochard-Wyart, M. Rakib and H. Duval, Soft Matter {\bf 7}, 96 (2011).
\bibitem{Yeomans}R. Ledesma-Aguilar, T. Sakaue and J.M. Yeomans, Soft Matter {\bf 8}, 1884 (2012); {\it ibid} {\bf 8}, 4306 (2012).

\bibitem{elastomeric_channel}D. Huh et. al., Nature. Mat. {\bf 6}, 424 (2007).
\bibitem{Annealed}A.M. Gutin, A.Y. Grosberg and E.I Shakhnovich, Macromolecules {\bf 26}, 1293 (1993).
\bibitem{Likos}A. Nikoubashman and C. Likos, J. Chem. Phys. {\bf 133}, 074901 (2010).
\bibitem{Hyperbranch}L. Li, C. He, W. He and C. Wu, Macromolecules {\bf 45}, 7583 (2012).




















\end{thebibliography}
\end{document}